\documentclass[aps,prl,reprint,pdftex,10pt,letterpaper,]{revtex4-1}
\usepackage{graphicx}
\usepackage{pslatex}
\usepackage[utf8]{inputenc}
\begin{document}

\title{Doping-dependent critical Cooper-pair momentum in thin, underdoped cuprate films}
\author{John Draskovic}
\email[]{draskovic.1@osu.edu}
\affiliation{Department of Physics, The Ohio State University, Columbus, Ohio, 43210, USA}
\author{Stanley Steers}
\affiliation{Department of Physics, The Ohio State University, Columbus, Ohio, 43210, USA}
\author{Thomas McJunkin}
\affiliation{Department of Physics, The Ohio State University, Columbus, Ohio, 43210, USA}
\author{Adam Ahmed}
\affiliation{Department of Physics, The Ohio State University, Columbus, Ohio, 43210, USA}
\author{Thomas R. Lemberger}
\affiliation{Department of Physics, The Ohio State University, Columbus, Ohio, 43210, USA}
\date{\today}

\begin{abstract}
We apply a recently-developed low-field technique to inductively measure the critical pair momentum $p_c$ in thin, underdoped films of Y$_{1-x}$Ca$_{x}$Ba$_{2}$Cu$_{3}$O$_{7-\delta}$  and Bi$_{2}$Sr$_{2}$CaCu$_{2}$O$_{8+\delta}$ reflecting a wide range of hole doping.  We observe that $p_c \propto \hbar/\xi$ scales with $T_c$ and therefore superfluid density $n_s(T\rightarrow0)$ in our two-dimensional cuprate films.  This relationship was famously predicted by a universal model of the cuprates with a \textit{doping-independent} superconducting gap, but has not been observed by high field measurements of the coherence length $\xi$ due to field-induced phenomena not included in the theory.
\end{abstract}

\maketitle

The phenomenology of underdoped, hole-doped cuprates is puzzling due to the observation of a multitude of characteristic spectroscopic energies, none of which appears to scale with the  superconducting critical temperature $T_c$.  On the other hand, there is significant evidence\cite{uemura_universal_1989},\cite{kondo_competition_2009} that $T_c$ is regulated by superfluid density $n_s$.  Previous work\cite{hetel_quantum_2007},\cite{hinton_comparison_2013} with thin (thickness $d <$ 5 unit cells) cuprate films shows that $T_c$ is consistent with an observed Kosterlitz-Thouless vortex-driven thermal phase transition and scales with $n_s(T\rightarrow0)$ as expected for a 2D quantum critical point at $n_s \rightarrow 0$. 

\par  Angle-Resolved PhotoEmission Spectroscopy (ARPES) reveals an  anisotropic gap structure in the cuprate superconductors, traditionally considered in a two-gap picture.  Plotted against the $d$-wave angular dependence $|cos(k_x a)-cos(k_y a)|/2$, the spectroscopic gap rises linearly in directions away from the node at ($\pi,\pi$) and bends sharply upward near the antinodal directions ($\pm \pi,0$) and (0,$\pm \pi$).  The gap magnitude near the node is seen to vanish for temperatures above the superconducting $T_c$ (for all but the lowest-doped samples), and a linear fit to the nodal gap\cite{yoshida_pseudogap_2011} gives a value $\Delta_0$.  The antinodal gap $\Delta^* > \Delta_0$ vanishes above the pseudogap temperature $T_c^*$.  It appears from data on underdoped samples\cite{yoshida_universal_2009} that $\Delta^* \approx 2 k_B T_c^*$ across the cuprates, rather like a dirty-limit BCS gap.  As the superconducting $T_c$ diminishes with underdoping, $T_c^*$ and the pseudogap $\Delta^*$ are seen to rise.

\par  On the overdoped side of the "superconducting dome" where the pseudogap is significantly depressed, $\Delta_0$ scales roughly with $T_c$, leading to its identification as the superconducting gap.  The situation is complicated on the underdoped side of the dome, where $\Delta_0$ remains \textit{constant} as the pseudogap emerges and $T_c$ falls.  The latest high-resolution Laser ARPES reveals that another gap opens at very low doping\cite{vishik_phase_2012} (hole concentration $p < 0.075$), with a nonzero magnitude at the node.  From these disparate data, a single spectroscopic energy that scales with $T_c$ is not obvious.  Based on the available ARPES data, Lee and Wen (L\&W) used a doping-independent $d$-wave gap in their generalized underdoped cuprate model\cite{lee_unusual_1997} and concluded that the superconducting coherence length $\xi$ is inversely proportional to $T_c$, rather than the energy gap, as in BCS superconductors.  Our two-coil experiments measure $p_c \propto \hbar/\xi$ directly, and thus make for a natural test of the theory.

\par Thin cuprate films were prepared via Pulsed Laser Deposition (PLD) using a 248-nm krypton-fluorine excimer laser. The laser was operated at a rate of 30 Hz with an energy density of 2.4 J/cm$^2$ at the target.  Growth proceeded in an atmosphere of flowing oxygen at 300 mTorr. The Y$_{1-x}$Ca$_{x}$Ba$_{2}$Cu$_{3}$O$_{7-\delta}$ (Ca-YBCO) films were grown on commercially manufactured strontium titanate (STO-001) substrates heated to 760$^{\circ}$ C, and the Bi$_{2}$Sr$_{2}$CaCu$_{2}$O$_{8+\delta}$ (Bi-2212) on lanthanum aluminum oxide (LAO-100) at 735$^{\circ}$ C.

\par Film thickness was controlled by measuring the pulses-per-unit-cell for a thick film using atomic force microscopy. STO substrates were prepared with three unit cells of (non-superconducting) PBCO deposited on the bare STO.  Ca-YBCO was deposited one unit cell at a time, with a 90 second rest between growth layers.  Bi-2212 was deposited one half unit-cell at a time directly to bare LAO, with an 80 second delay between layers.  All film samples were then capped with ten unit cells of PBCO to protect against atmospheric degradation.  After deposition, the Ca-YBCO films were annealed at 450$^{\circ}$ C in oxygen pressures ranging from 1 to 650 Torr to control the oxygen doping.  The Bi-2212 films were annealed at 700$^{\circ}$ C in pressures of 75 to 100 Torr.  The hole concentration is not measured, so the primary measure of doping is the superfluid density $n_s(T\rightarrow0)$.

\par  Thin film measurements were conducted in a liquid helium cryostat capable of 1.4 K.  Superfluid density was measured in a typical two-coil experiment\cite{turneaure_numerical_1996},\cite{turneaure_numerical_1998}, using a small 10 kHz drive current and a correspondingly small drive field ($<20$ mG).  Throughout the paper we shall refer to the quantity $1/\lambda^2(T)$, proportional to the three dimensional density of supercurrent carriers, as the "superfluid density".  The \textit{areal} superfluid density $n_s d$ is proportional to $1/\Lambda$, where $\Lambda \equiv 2\lambda^2/d$ is the two-dimensional penetration depth called the Pearl length.  The two-coil experiment measures $1/\Lambda$ directly.  Values of $1/\lambda^2$ therefore contain any uncertainty in the film thickness.

\begin{figure}
\begin{center}
	\includegraphics[width=0.4\textwidth]{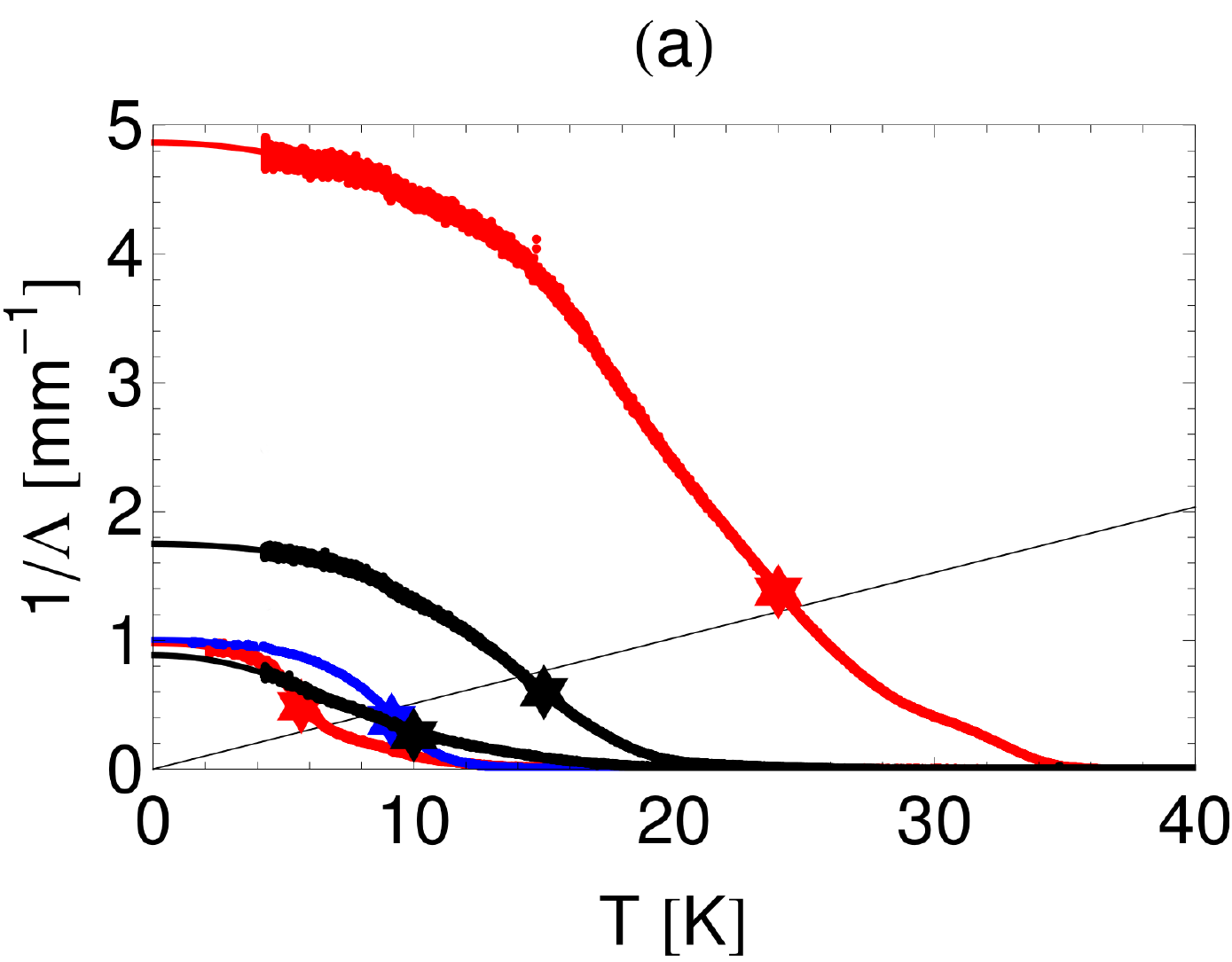}\\
	\includegraphics[width=0.4\textwidth]{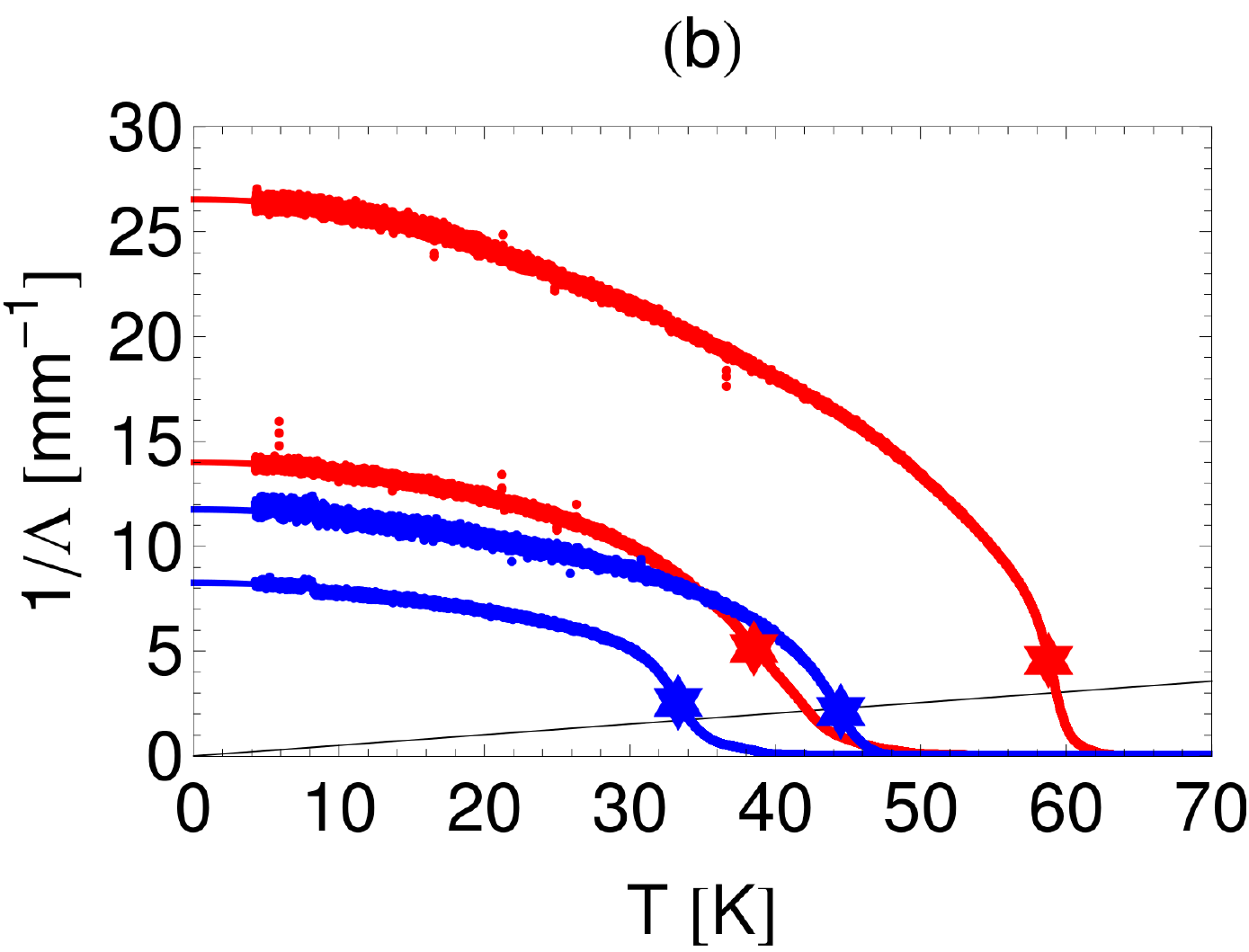}\\
	\includegraphics[width=0.4\textwidth]{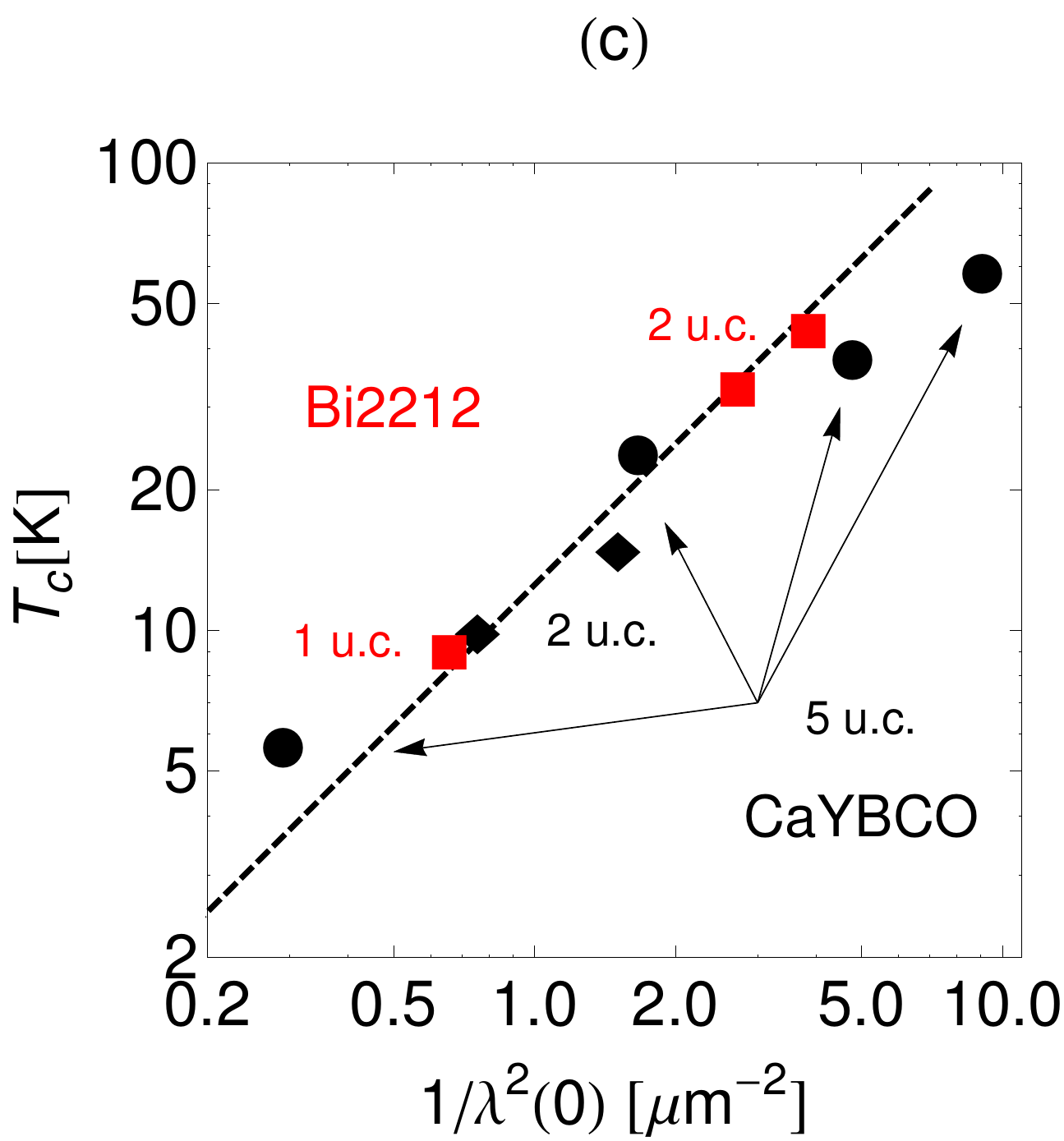}\\
\end{center}
	\caption{Areal superfluid density $1/\Lambda(T)$ measured for the nine films in this study: five-unit-cell Ca-YBCO  (red: a,b), two-unit-cell Ca-YBCO (black: a), one-unit-cell Bi2212 (blue: a), and two-unit-cell Bi2212 (blue: b).  Lowest-doping films exhibit a downturn in superfluid predicted by a Kosterlitz-Thouless (KT) transition: $1/\Lambda(T) = 4 \pi \mu_0 k_B T/\phi_0^2$, assuming the full film thickness acts as a 2D unit.  $T_c$ (stars: a,b) is defined by the peak in the real part of the measured conductivity (\textit{not shown}), coincident with the predicted KT phase transition, and scales as expected for two-dimensional quantum critical fluctuations (c).  Dashed line is the 2D scaling reported by  Hetel et al.\cite{hetel_quantum_2007}.}
	\label{fig:superfluid}
\end{figure}

\par Figures \ref{fig:superfluid}a, \ref{fig:superfluid}b, and \ref{fig:superfluid}c give the superfluid density as a function of temperature for the nine films in this study.  We analyzed the data in the manner of Hetel et al.\cite{hetel_quantum_2007}, and the scaling of the extrapolated zero-temperature superfluid density $1/\lambda^2(0)$ with the transition temperature in Figure \ref{fig:superfluid}d follows that of Ref. \cite{hetel_quantum_2007}, as well as Hinton et al.\cite{hinton_comparison_2013}, so we conclude that the present films are comparable.

\begin{figure}
\begin{center}
 \includegraphics[width=0.4\textwidth]{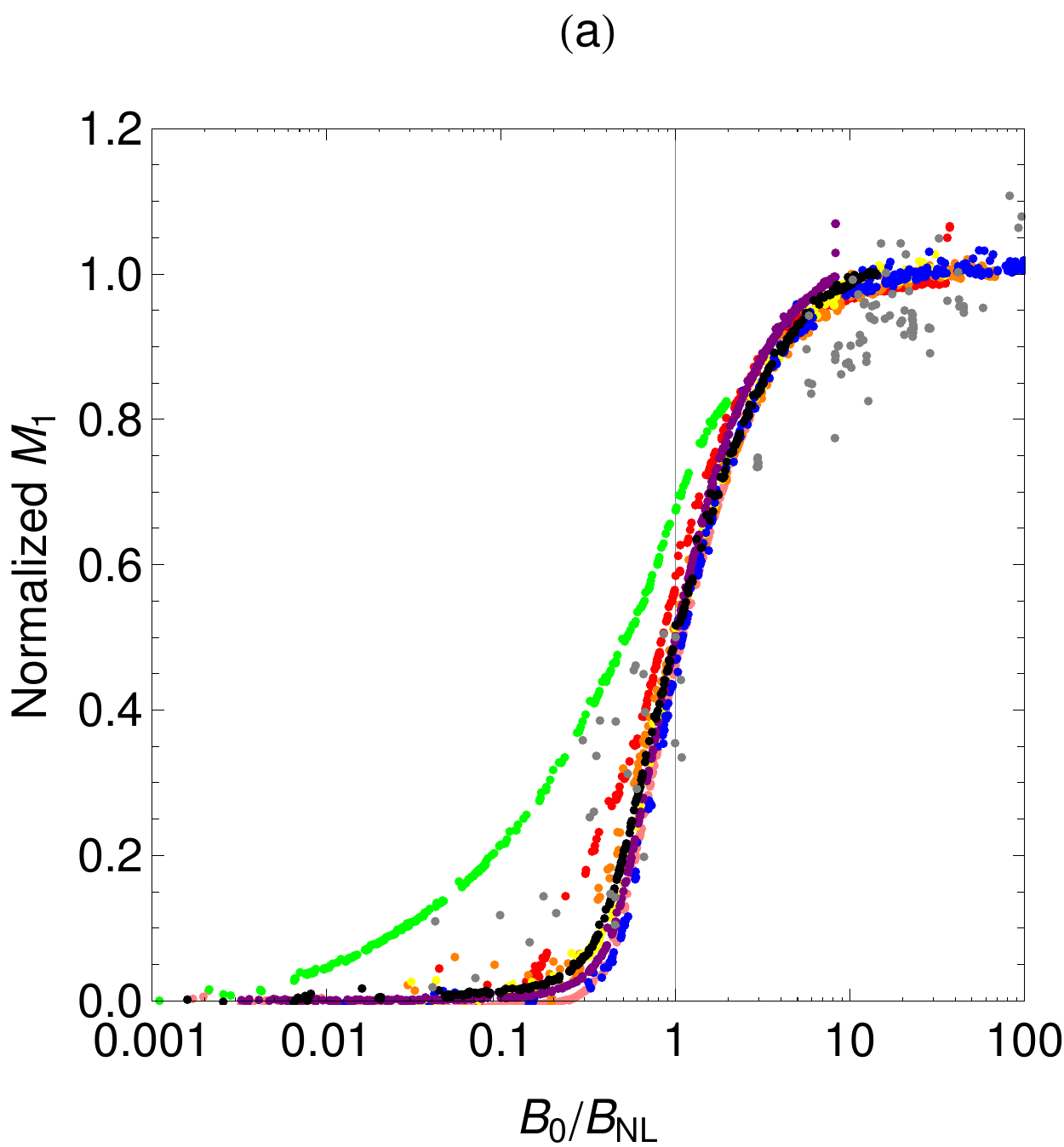} \\ 
\includegraphics[width=0.4\textwidth]{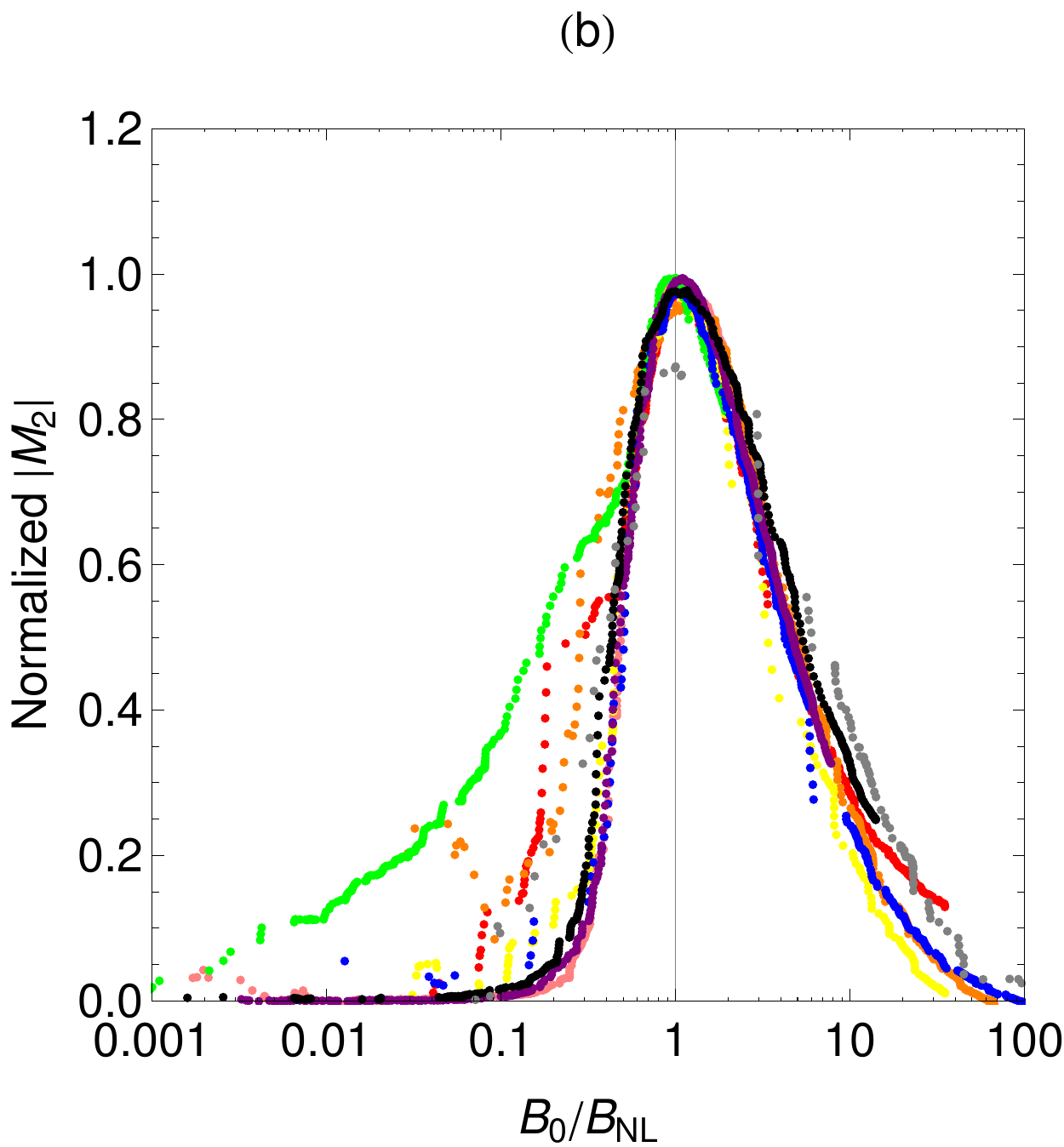} \\
 \end{center} 
\caption{In-phase ($M_1$) and out-of-phase ($M_2$) mutual inductances as functions of drive field at the center of the film ($B_0$) for all nine films in the study (at 10 kHz frequency).  Normalization is as follows: $M_1 = \frac{M_1(B_0)-M_1(B\rightarrow0)}{M_0-M_1(B\rightarrow0)}, M_2 = \frac{|M_2(B_0)|}{\textrm{max}|M_2(B_0)|}$.  Signal is phase-calibrated at low excitation field $B_0\rightarrow0$: $M_2 \rightarrow 0$.  Initial position is measured from the normal-state response: $M_0 \equiv M_1(T>T_c)$.}
\label{fig:nonlinear}
\end{figure}

\par The samples were transferred to a second cryostat, where they were again cooled as low as 1.4 K to measure the nonlinear response\cite{draskovic_measuring_2013}.  Our superconducting drive coil is designed so that the maximum amplitude $B_0$ of the applied magnetic field is perpendicular to the film surface at the center of the film.  $B_0$ is smoothly varied from 0-40 Gauss at constant temperature to produce the mutual inductance traces presented in Figure \ref{fig:nonlinear}.  The normalization reveals the distinct peak feature in the out-of-phase mutual inductance common to all of the films, regardless of their thickness and doping, and this peak tracks the field at which the in-phase mutual inductance rises to midway between the normal and superconducting state values.  We identify this value of $B_0$ as $B_{NL}$, the field at which the induced supercurrent density in the film reaches its critical value and vortex-antivortex pairs unbind as the metastable Meissner state breaks down.  This interpretation of the data was verified qualitatively and quantitatively from measurements on thin Nb and MoGe films\cite{draskovic_measuring_2013}.

\begin{figure}
	\begin{center}
		\includegraphics[width=0.4\textwidth]{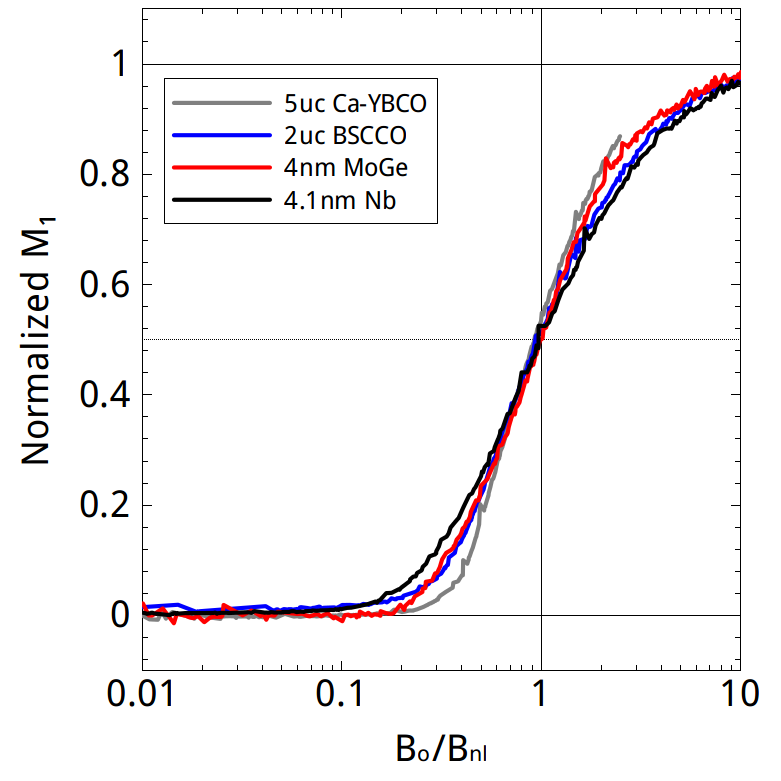}
		\includegraphics[width=0.4\textwidth]{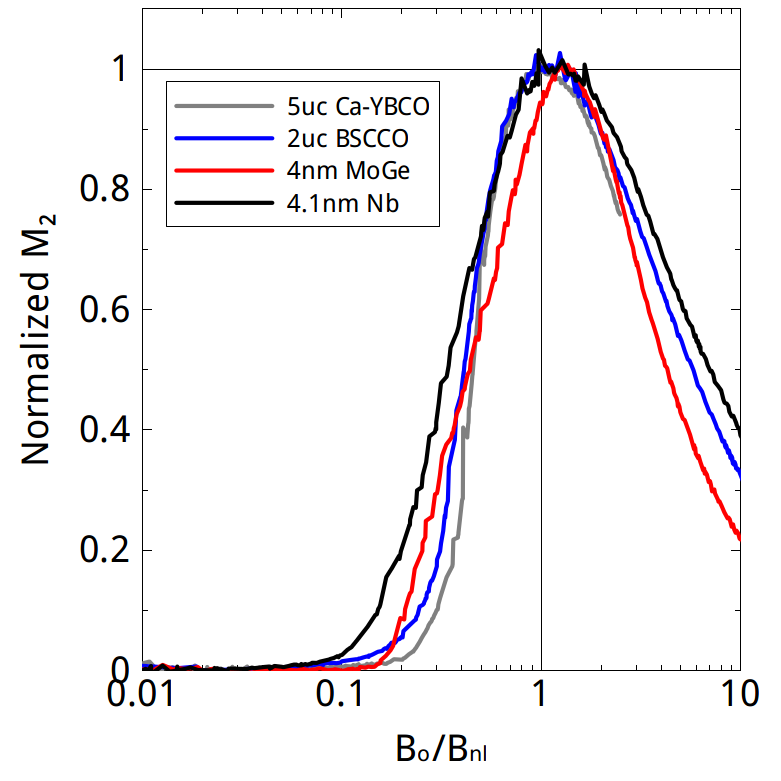}
	\end{center}
\caption{Comparison of nonlinear two-coil response for four materials.  Normalization follows Fig. \ref{fig:nonlinear}.}
	\label{fig:similar}
\end{figure}

\par
\begin{figure}
	\begin{center}
		\includegraphics[width=0.4\textwidth]{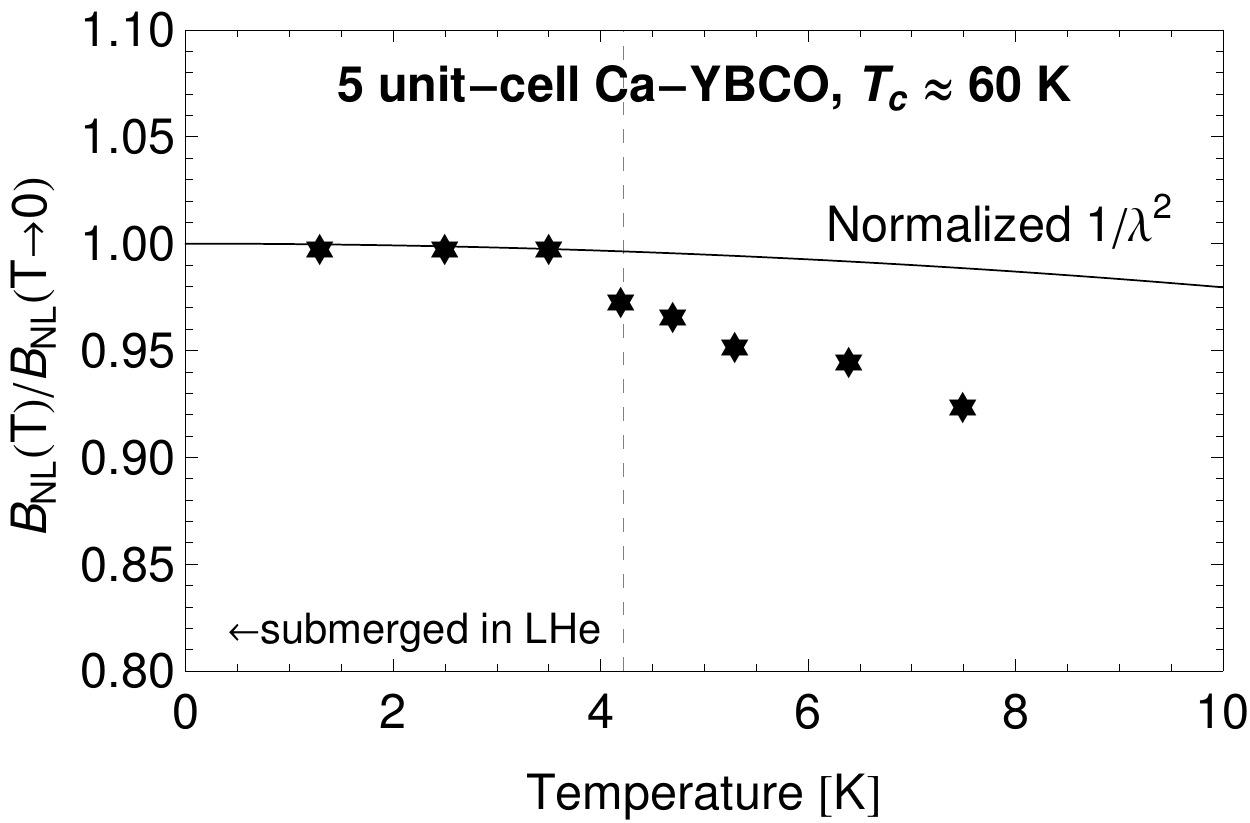}
	\end{center}
	\caption{$B_{NL}$ measured at fixed temperatures for the $T_c \approx $ 60 K film. Below 4.2 K, the experiment is submerged in liquid helium and $B_{NL}$ is constant.  Above 4.2 K, $B_{NL}$ falls slightly due to heating.  Heating is a 10\% effect up to 8 K.}
	\label{fig:tempdep}
\end{figure}

  The critical pair momentum $p_c$ is computed from the low-temperature values of $B_{NL}$ by a calculation of Lemberger and Ahmed\cite{lemberger_upper_2013} using an idealized version of our experimental configuration.  We report values of $p_c$ for our cuprate films using this result, assuming that the measured $B_{NL}$ coincides with the applied field at which the Meissner state is becomes unstable, and vortex-antivortex pairs unbind \textit{en masse}\cite{lemberger_theory_2013}.  For $T < 8$ K, $B_{NL}$ is essentially flat (Figure \ref{fig:tempdep}), consistent with calculations of $p_c(T)$ in s-wave\cite{romijn_critical_1982} and d-wave superconductors\cite{khavkine_supercurrent_2004}.  This is in sharp contrast to the measured vortex lattice-melting field\cite{ramshaw_vortex_2012}, which has a large negative slope near $T \rightarrow 0$.

\par   We first note the universal qualitative behavior of the superfluid density depicted in Figure \ref{fig:superfluid}a and b, where various CaYBCO films and Bi2212 films are plotted in a thickness-independent manner.  The step-down in superfluid near the predicted Kosterlitz-Thouless transition temperature is common to all films.  Our single- and double-unit-cell Bi2212 films, which are considerably thinner than those studied by Yong et al.\cite{yong_evidence_2012}, do not show the linear suppression of superfluid with temperature observed by those authors in samples with $T_c <$ 50 K.  As the present films follow the two-dimensional scaling of $T_c$ with $n_s$ observed by Ref. \cite{hetel_quantum_2007}, we conclude that our 1-5 unit-cell-thick films are effectively two-dimensional, with superfluid density suppressing $T_c$ in the vicinity of a quantum critical point at $T_c \rightarrow 0$.

\par  Next, the nonlinear mutual inductance traces in Figure \ref{fig:nonlinear} are noteworthy in their self-consistency, as well as their qualitative consistency with the BCS superconductors Nb and MoGe (Fig. \ref{fig:similar}), studied previously\cite{draskovic_measuring_2013}.  The similarity between the nonlinear traces in dirty-limit BCS films and 2D cuprate films supports the interpretation of these data as a general phenomenon of superconducting films.

\begin{figure}
	\begin{center}
		\includegraphics[width=0.4\textwidth]{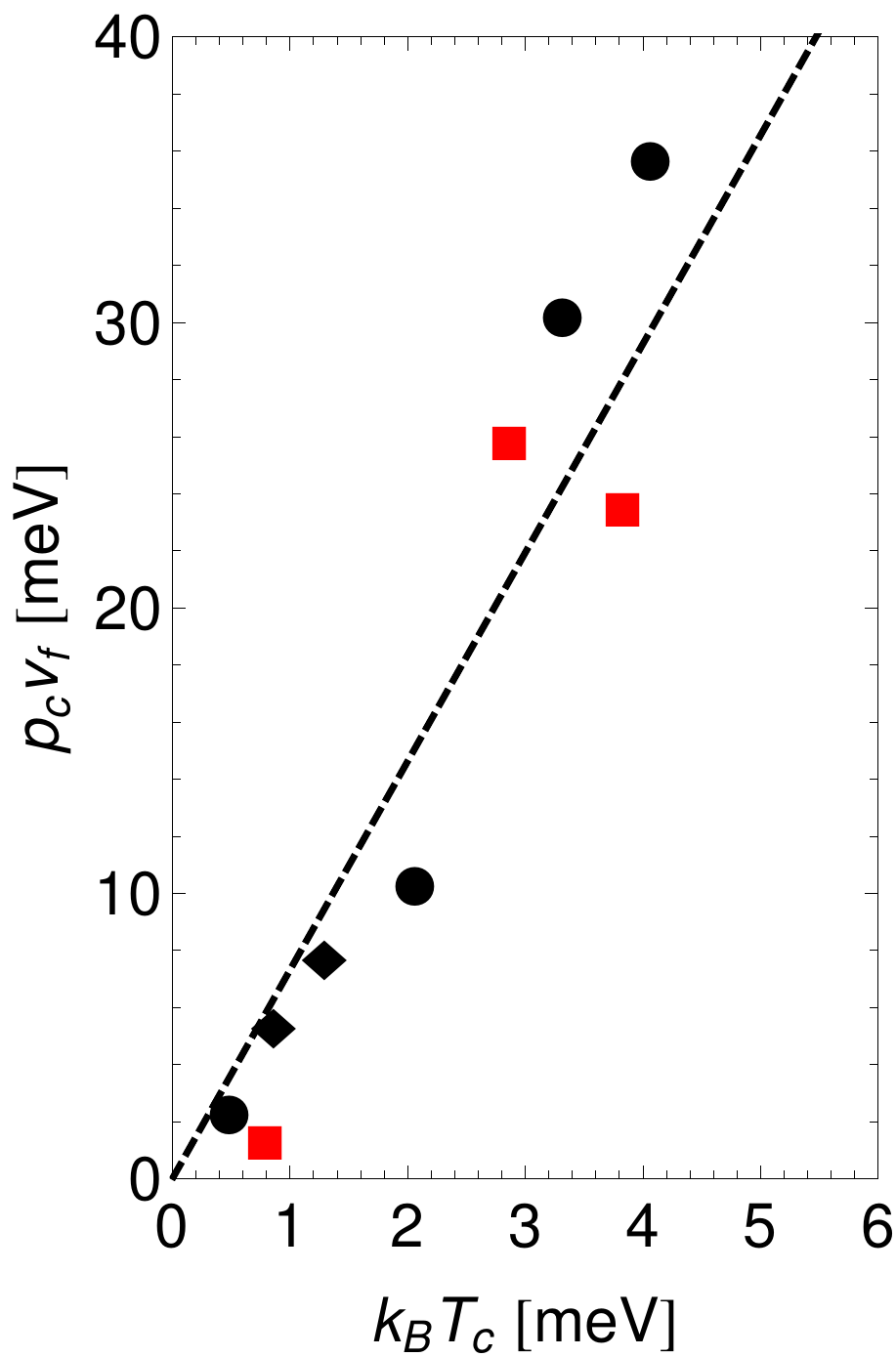}
	\end{center}
		\caption{Measured critical pair momentum $p_s$ plotted against $T_c$.  Plotting conventions follow Fig. \ref{fig:superfluid}c.  $v_f = 3 \times 10^5$ m/s, the universal nodal Fermi velocity measured by ARPES\cite{zhou_high-temperature_2003}.  Dashed line is empirical linear fit.}
		\label{fig:pc}
\end{figure}

\par  Finally, Figure \ref{fig:pc} is the major result of the present study: we observe that the critical pair momentum scales with the measured $T_c$ in thin, underdoped cuprate films.  For an s-wave BCS superconductor, this is expected because $T_c$ gives a measure of the energy gap, and depairing\cite{bardeen_critical_1962} occurs when the applied vector potential un-gaps the quasiparticle spectrum.  In the cuprates, however, the gap is seemingly doping-independent, so it is not obvious that $p_c$ \textit{should} scale with $T_c$.  Nonetheless, taking $p_c$ as inversely proportional to L\&W's coherence length we expect to see $p_c \propto T_c$ as observed in our data.

The latest high-magnetic field measurements\cite{grissonnanche_direct_2014} of $B_{c2} \propto 1/\xi^2$ affirm a non-monotonic increase of $B_{c2}$ with $T_c$, rather than the simple quadratic dependence that follows from the L\&W model.  Given that spin order and charge order have been directly observed in conjunction with the vortex lattice\cite{hoffman_four_2002},\cite{lake_antiferromagnetic_2002},\cite{wu_magnetic-field-induced_2011} and the vortex radius itself is seen to be field-dependent\cite{sonier_expansion_1999}, it is not surprising that the L\&W theory does not fit the $B_{c2}$ data.  The framework of L\&W simply does not include such physics, which appear to become important at \textit{much} higher applied magnetic fields than our experiment.


\par In summary, we performed an inductive measurement of the critical pair momentum in nine thin, underdoped, hole-doped cuprate films.  $T_c$ is proportional the zero-temperature superfluid density $n_s(T\rightarrow 0)$, consistent with two-dimensional quantum critical behavior.  The critical pair momentum is observed to scale with $T_c$ and thus $n_s(T\rightarrow 0)$, directly confirming the theory of Lee and Wen\cite{lee_unusual_1997}, which assumes a doping-independent superconducting gap consistent with the latest ARPES studies, where no evidence of a diminishing gap as $T_c \rightarrow 0$ is observed.

\section*{Acknowledgments}

\par The authors wish to thank Mohit Randeria and Sumilan Banerjee for their interest and input.  This research was supported by DOE-Basic Energy Sciences through Grant No. FG02-08ER46533.

\bibliographystyle{apsrev4-1}
\bibliography{draft9}

\end{document}